\documentclass[11pt,a4paper]{article} 

\usepackage{jcappub} 

\usepackage{graphicx}
\usepackage{dcolumn}
\usepackage{bm}
\usepackage[english]{babel}
\usepackage[applemac]{inputenc} 
\usepackage{graphicx}
\usepackage{epstopdf}
\usepackage{amsfonts}
\usepackage{color}
\usepackage{epsfig}
\usepackage{mathrsfs}
\usepackage{amsmath,amssymb}
\usepackage{subfigure}

\usepackage{framed}
\usepackage{lipsum}
\usepackage[svgnames]{xcolor}
\definecolor{shadecolor}{named}{LightGrey}

\newcommand{\lto}[1]{\longrightarrow#1}

\renewcommand{\(}{\left(}
\renewcommand{\)}{\right)}

\begin{document}

\graphicspath{{figure/}}
\selectlanguage{english}

\title{Fast generation of mock maps from realistic catalogs of astrophysical objects}

\author[a]{M. De Domenico}
\author[b]{H. Lyberis}

\affiliation[a]{School of Computer Sciences, University of Birmingham, Birmingham, UK}
\affiliation[b]{Universidade Federal do Rio de Janeiro, Instituto de Fisica, Rio de Janeiro, RJ, Brazil}

\emailAdd{m.dedomenico@cs.bham.ac.uk}
\emailAdd{lyberis.haris@gmail.com}

\date{\today}

\abstract{
We present a novel method to generate a synthetic distribution of objects (mock) on a spherical surface (i.e. a sky), by using a real distribution. The resulting surrogate map mimics the clustering features of the real data, including the effects of non-uniform exposure, if any.
The method is model-independent, also preserving the angular correlation function, as well as the angular power spectrum, of the original data. It can be reliably adopted to mimic the angular clustering of objects distributed on a spherical surface and it can be easily extended to include further information, as the spatial clustering of objects distributed inside a sphere. Applications to real data are presented and discussed. In particular, we consider the distribution of galaxies recently presented in the 2MASS Redshift Survey (2MRS) catalog.
}


\notoc

\maketitle

\flushbottom


\section{Introduction}

Redshift surveys reveal that the distribution of matter in the Universe appears to follow patterns typical of strongly correlated stochastic processes, from smallest to largest scales \cite{Peebles-1980,Davis-1983,ramella19902ptcorfunc}. Numerical simulations of nonlinear gravitational clustering in Universes dominated by weakly interacting cold dark matter have been shown to nicely reproduce the salient features of the observed clustering, as well as its formation and evolution \cite{davis1985evolution}. Recently, simulations of the growth of dark matter structure, from high redshifts to the present time, have shown a considerable agreement with current models for the formation of structure in the Universe \cite{springel2005simulations}. Such simulations are able to provide very realistic distributions of matter in the Universe.

During the last decades the search for statistically significant clustering of such a matter has played a crucial role for understanding the cosmological structure formation. A standard approach involves the estimation of the statistical autocorrelation function of the matter density field. For such a purpose, many estimators have been proposed and widely used to capture clustering structures by measuring the deviation from isotropy of angular or radial distributions \cite{Davis-1983, Szalay-1993, Hamilton-1993}. 

Under the assumption that matter is randomly distributed in a space $\mathcal{V}$ according to a Poisson process, the number of objects at a distance $r$ is given by
\begin{eqnarray}
dN_{12} = \bar{n}^{2}dV_{1}dV_{2},
\end{eqnarray}
where $\bar{n}$ is the expected number density of objects in $\mathcal{V}$, $dV_{1}$ and $dV_{2}$ are two volume elements. However, if the underlying distribution is not Poissonian, the density field is subjected to fluctuations inducing clustering of objects. Such an inhomogeneity can be captured by inspecting the two-point autocorrelation function $\xi_{12}(r)$ (2ptACF), defined by
\begin{eqnarray}
dN_{12} = \bar{n}^{2}\(1+\xi_{12}(r)\)dV_{1}dV_{2}.
\end{eqnarray}
If $dN_{12}$ is normalized to the whole volume of the space $\mathcal{V}$, the above definition suggests that the 2ptACF provides a measure of the probability to find an excess of objects at the scale $r$. 

Within the present work, we show how to generate random realizations (\emph{i.e.} surrogates) of a true distribution of objects, which preserve the angular clustering features of the original data. Our method is rather general: it can be extended to higher dimensions and it is suitable for many different applications. In particular, we consider the case of astrophysical and cosmological studies, where it can be used to build catalogs of objects which mimic the clustering features of real matter in the Universe. For instance, such surrogate catalogs can be used to train new or existing methods for clustering detection~\cite{Peebles-1980,ave20092pt+,Hague09,Gaztanaga94Clusteringmethod,dedomenico2011MAF}, and they are suitable to simulate realistic distributions of sources of ultra-high energy cosmic rays, in anisotropy or source density studies~\cite{cuoco2008clustering,cuoco2009global,DeDomenicoICRC2011,takami2012sourcedensitylimit}.

The paper is organized as follows: in Sec.\,\ref{sec:cluster} we briefly describe the standard correlation function adopted to investigate the clustering of astrophysical objects and we show its relationship with the corresponding power spectrum. In Sec.\,\ref{sec:surrogate} we present our method to produce surrogate catalogs, based on phase randomization in the spherical harmonics space. We will show that the angular power spectrum of the surrogates matches that of the original catalog, and, by consequence, that the corresponding correlation function is preserved. We consider the application to objects synthetically distributed on the sphere and to real-world objects, namely the galaxies in the 2MASS Redshift Survey catalog (2MRS) \cite{2MRScatalog2011}.


\section{Clustering identification}\label{sec:cluster}

\subsection{The correlation function}

If $\rho(\mathbf{r})$ denotes the density field as a function of the position $\mathbf{r}$, and 
\begin{eqnarray}
\label{def-field-perturb}
\delta(\mathbf{r})=\frac{\rho(\mathbf{r})-\langle\rho(\mathbf{r})\rangle}{\langle\rho(\mathbf{r})\rangle}
\end{eqnarray}
indicates the density field perturbation, the standard definition of the statistical 2ptACF is given by
\begin{eqnarray}
\label{def-autocorr}
\xi(\mathbf{r})=\langle\delta(\mathbf{x})\delta(\mathbf{x}+\mathbf{r})\rangle,
\end{eqnarray}
$\langle\cdot\rangle$ denoting an averaging over the normalization volume $V$ \cite{peacock2003large}. More general measures, assuming an underlying fractal structure of the matter distribution, make use of modified correlation functions based on an average density depending on the position and on the size of the sample volume considered, in the framework of fractal analysis \cite{pietronero1987fractal,coleman1992fractal,martinez1999Universe,joyce2000fractal}.

Because of the lack of large redshift samples, the first estimations of $\xi(\mathbf{r})$ has been given by the two-point angular correlation function $\omega(\theta)$, providing a weighted projection of $\xi(\mathbf{r})$ on a two-dimensional sky through the Limber equation \cite{Peebles-1980}
\begin{eqnarray}
\omega(\theta)=\int_{0}^{\infty}dy\,y^{4}\phi^{2}\int_{-\infty}^{\infty}dx\,\xi\(\sqrt{x^{2}+y^{2}\theta^{2}}\),
\end{eqnarray}
where $\phi$ is the radial selection function for the considered sky survey.

Hence, the problem of estimating $\xi(\mathbf{r})$ in a three-dimensional space has been reduced to the problem of estimating $\omega(\theta)$ on a two-dimensional sky. If $n$ is the number of objects on the projected spherical surface $\mathcal{S}$ of $\mathcal{V}$ and $m$ is the number of objects coming from several Monte Carlo realizations of $\mathcal{S}$, the angular autocorrelation function can be simply estimated by looking for excesses of objects, at a given angular scale, with respect to the isotropic expectation. Many estimators have been introduced for such a purpose, whose differences are mainly related to the biased estimation they provide. For sake of completeness, we mention here the most known: Peebles \cite{Peebles-1980}, Davis-Peebles, Landy-Szalay and Hamilton \cite{Davis-1983, Szalay-1993, Hamilton-1993} angular correlation functions, respectively defined as
\begin{eqnarray}
\label{def-w1}
\omega_{\text{P}}(\theta) &=& \frac{m(m-1)}{n(n-1)}\frac{DD(\theta)}{RR(\theta)}-1,\\
\label{def-w2}
\omega_{\text{DP}}(\theta) &=& \frac{2m}{n-1}\frac{DD(\theta)}{DR(\theta)}-1,\\
\label{def-w3}
\omega_{\text{LS}}(\theta) &=& \frac{m(m-1)}{n(n-1)}\frac{DD(\theta)}{RR(\theta)}-\frac{m-1}{n}\frac{DR(\theta)}{RR(\theta)}+1,\\
\label{def-w4}
\omega_{\text{H}}(\theta) &=& \frac{4nm}{(n-1)(m-1)}\frac{DD(\theta)\times RR(\theta)}{DR^{2}(\theta)}-1,
\end{eqnarray}
where $DD$ is the number of pairs lying between $\theta$ and $\theta+\Delta \theta$ for the experimental distribution, $RR$ is the same number calculated for Monte Carlo realizations and $DR$ is the cross-pair counts between experimental and simulated events. By definition (see Ref. \cite{Peebles-1980}) the functions (\ref{def-w1})-(\ref{def-w4}) are strictly related to the excess on the area $d\Omega$ of the pairs number $dN_{\text{pairs}}$ over the random background $dN_{\text{MC}}$:
\begin{eqnarray}
dN_{\text{pairs}}-dN_{\text{MC}}=\frac{1}{2}nn_{0}\omega(\theta)d\Omega,
\end{eqnarray}
where $n_{0}$ is the expected average density of events on $\mathcal{S}$ assuming an isotropic distribution.

The large number of objects in the sky surveys, from the earliest to the latest ones, have ensured a good estimation of $\omega(\theta)$, further reducing the error on the more recent estimates. On small angular scales a power-law behaviour 
\begin{eqnarray}
\omega(\theta)=\(\frac{\theta}{\theta_{0}}\)^{1-\gamma}
\end{eqnarray}
has been observed, where $\gamma\approx 1.8$ \cite{Peebles-1980,guzzo1997redshift} is the index of the spatial autocorrelation counterpart 
\begin{eqnarray}
\xi(r)=\(\frac{r}{r_{0}}\)^{\gamma},
\end{eqnarray}
where $\mathbf{r}$ has been replaced by $r$, namely the distance between two objects in the space, and $r_{0}\approx 5h^{-1}$ Mpc, generally indicated as the correlation length, is the distance such that the clustering strength does not exceed the homogeneous expectation. The exponent $\gamma$ is somehow universal: it is the same for galaxies, cluster of galaxies and superclusters in the Universe. However, the correlation function shows a sudden break around $r>10h^{-1}$ Mpc, above which $\xi(r)$ rapidly approaches zero. Moreover, it is worth remarking that a direct estimation of $\xi(r)$ is difficult because of the effects induced by galaxy peculiar velocities, directly related to the evolution of the density field through conservation of mass \cite{peacock1999cosmological,peacock2001measurement,peacock2003large,hawkins20032df}. Such effects induce $\xi(r)$ to deviate from a power-law, mainly because separations are estimated in the redshift space, instead of the real three-dimensional space \cite{guzzo1997redshift}. The direct consequence is a flatter correlation function than $\xi(r)$ in the real-space, although, fortunately, reliable corrections exist \cite{fisher1994clustering,ghigna1996deviations,saunders2000density}.

\subsection{The power spectrum}

Let us suppose that the field $\delta(\mathbf{r})$, defined by Eq.\,(\ref{def-field-perturb}), is periodic within a box of size $L$ and let us consider the correlation function of Eq.\,(\ref{def-autocorr}). By expanding the density field perturbation $\delta(\mathbf{r})$ in its Fourier modes, we obtain
\begin{eqnarray}
\xi(\mathbf{r})&=&\left\langle \(\sum_{\mathbf{k}}\delta_{\mathbf{k}}e^{i\mathbf{k}\cdot\mathbf{x}}\)\(\sum_{\mathbf{k'}}\delta_{\mathbf{k'}}e^{i\mathbf{k'}\cdot(\mathbf{x}+\mathbf{r})}\) \right\rangle=\left\langle \sum_{\mathbf{k}}\delta_{\mathbf{k}}\delta_{\mathbf{-k}}e^{-i\mathbf{k}\cdot\mathbf{r}} \right\rangle,\nonumber
\end{eqnarray}
where, because of the periodic boundary conditions $k_{i}=n2\pi/L$ ($i=x,y,z$), the cross terms with $\mathbf{k}+\mathbf{k'}\neq 0$ average to zero. Indeed, by taking into account that $\xi(\mathbf{r})$ is a real function, it follows $\delta_{-\mathbf{k}}=\delta_{\mathbf{k}}^{*}$ and
\begin{eqnarray}
\xi(\mathbf{r})&=&\left\langle \sum_{\mathbf{k}}|\delta_{\mathbf{k}}|^{2}e^{-i\mathbf{k}\cdot\mathbf{r}} \right\rangle.\nonumber
\end{eqnarray}
Hence, by allowing the box size to become arbitrarely large, we can rewrite the inner sum as an integral in the $\mathbf{k}-$space and we can express $\xi(\mathbf{r})$ by volume averaging the resulting expression as
\begin{eqnarray}
\label{def-autocorr-spec}\xi(\mathbf{r})&=&\frac{V}{(2\pi)^{3}} \int|\delta_{\mathbf{k}}|^{2}e^{-i\mathbf{k}\cdot\mathbf{r}} d\mathbf{k},
\end{eqnarray}
where $\mathcal{P}(\mathbf{k})=\langle|\delta_{\mathbf{k}}|^{2}\rangle$ is the power spectrum of $\delta(\mathbf{x})$, quantifying the scale-dependence of density perturbations in the Fourier space \cite{peacock1999cosmological}. Eq.\,(\ref{def-autocorr-spec}) is commonly known as the Wiener-Khinchin-Kolmogorov theorem, stating that the autocorrelation function is the Fourier transform of the power spectrum, or, equivalently, that $\xi(\mathbf{r})$ and $\mathcal{P}(\mathbf{k})$ are Fourier pairs.

By volume averaging, we have implicitly assumed the ergodicity of the stochastic process underlying the field $\delta(\mathbf{r})$: in practice, this requirement is only supposed to be satisfied if a sufficiently large volume is considered \cite{peacock1999cosmological}.




\section{Surrogate distributions}\label{sec:surrogate}

In this section we present a method to generate a surrogate catalog of objects mimicking the clustering features of real catalogs of objects. Because of the Wiener-Khinchin-Kolmogorov theorem, two different skies of objects with the same angular (spatial) power spectrum have the same angular (spatial) correlation function. Therefore, any distribution of objects preserving the power spectrum $\mathcal{P}_{\text{data}}$ will preserve the clustering features of the real distribution. In the following, we will present a method which preserves such an angular power spectrum by means of a phase randomization approach in the Fourier space.

In the particular case of the angular distribution on the sphere, the formalism of spherical harmonics can be adopted. Although our approach is equivalent in the case of a multi-dimensional distribution, in the following we will focus on the two-dimensional case.

Let $(\theta,\phi)$ indicate a direction on the unit sphere. Moreover, at this step we assume a uniform exposure, whereas the case of a non-uniform exposure will be considered further in the text. Any square-integrable function $f(\theta,\phi)$ representing, for instance, an intensity map, can be expanded as a linear combination:
\begin{eqnarray}
\label{def-sht}f(\theta,\phi)=\sum_{\ell=0}^{\infty}\sum_{m=-\ell}^{\ell}a_{\ell m}Y_{\ell m}(\theta,\phi)
\end{eqnarray}
where $a_{\ell m}$ are the multipole coefficients and the functions 
\begin{eqnarray}
Y_{\ell m}(\theta,\phi)=\sqrt{\frac{(2\ell+1)}{4\pi}\frac{(\ell-m)!}{(\ell+m)!} P_{\ell m}(\cos\theta)e^{im\phi} }
\end{eqnarray}
represent the spherical harmonics, $P_{\ell m}$ indicates the Legendre polynomials and $\ell$ indicates the multipole moment. It is straightforward to show that spherical harmonics forms a complete set of orthonormal functions such that
\begin{eqnarray}
\int_{0}^{\pi}\int_{0}^{2\pi}Y_{\ell m}(\theta,\phi)Y_{\ell' m'}(\theta,\phi)\sin\theta d\phi d\theta=\delta_{\ell\ell'}\delta_{mm'},\label{eq:orthogonqlity}
\end{eqnarray}
being $\delta_{ij}$ the Kronecker delta. Finally, the observed angular power spectrum for the intensity map is defined by
\begin{eqnarray}
C_{\ell}=\frac{1}{2\ell+1}\sum_{m=-\ell}^{\ell}|a_{\ell m}|^{2}, 
\end{eqnarray}
as a function of the multipole moment $\ell$. The investigation of $C_{\ell}$ provides the same amount of information about clustering and anisotropy obtained from the investigation of $\mathcal{P}(\mathbf{k})$, defined at the end of the previous section. The angular correlation function can be obtained from the Fourier transform of $C_{\ell}$ on the unit sphere.

Our method is based on the fact that any change of the phases $\phi$, in the coefficients $a_{\ell m}$, does not affect neither the magnitudes nor the angular power spectrum $C_{\ell}$. Quantitatively, $C_{\ell}$ is invariant under the transformation $a_{\ell m}\lto a_{\ell m}e^{i\varphi_{\ell m}}$, where $\varphi_{\ell m}$ is a real number. The dependence on $\ell$ and $m$ is stressed because, in principle, a different phase can be associated to each multipole coefficient.

The method of phase randomization requires that $\varphi_{\ell m}$ is sampled from a stochastic variable, uniformly distributed in the interval between 0 and $2\pi$. Hence, the new multipole coefficients $a'_{\ell m}$ define a new intensity map $f'(\theta,\phi)$, while the corresponding sky provides a new random realization of the data, which preserves the clustering features. 
We choose to use HEALPix~\cite{HEALPix2005} to perform the discrete spherical harmonic transform of the skies. 
In the following, we indicate with $\mathcal{S}$ the spherical surface where the whole sky is defined. In general, it may happen that only a fraction $\mathcal{S}_{0}$ of $\mathcal{S}$ can be physically observed or it can be of interest for some reason. For a given pixelization of the sphere, we introduce a weight function, along any direction $(\theta,\phi)$ on the unit sphere, defined by
\begin{eqnarray}
\omega (\theta,\phi)&\propto&\left\{
\begin{array}{ll}
\Omega_{p}^{-1} & \text{if } (\theta,\phi)\in \mathcal{S}_{0} \label{eqn:windowfunc}\\
0 & \text{otherwise},
\end{array}\right.\nonumber
\end{eqnarray}
which assigns a weight proportional to $\Omega_{p}^{-1}$ to the pixel corresponding to the direction $(\theta,\phi)$ of an object, being $\Omega_{p}$ the solid angle covered by that pixel. Hence, the coefficients in the expansion given by Eq.\,(\ref{def-sht}) include the information provided by the weighting procedure. In the following, we distinguish two cases: i) sky with uniform coverage, i.e. $\mathcal{S}_{0}=\mathcal{S}$, and ii) sky with non-uniform coverage, i.e. $\mathcal{S}_{0}\subset\mathcal{S}$.

\begin{figure}[!t]
\includegraphics[width=8.0cm]{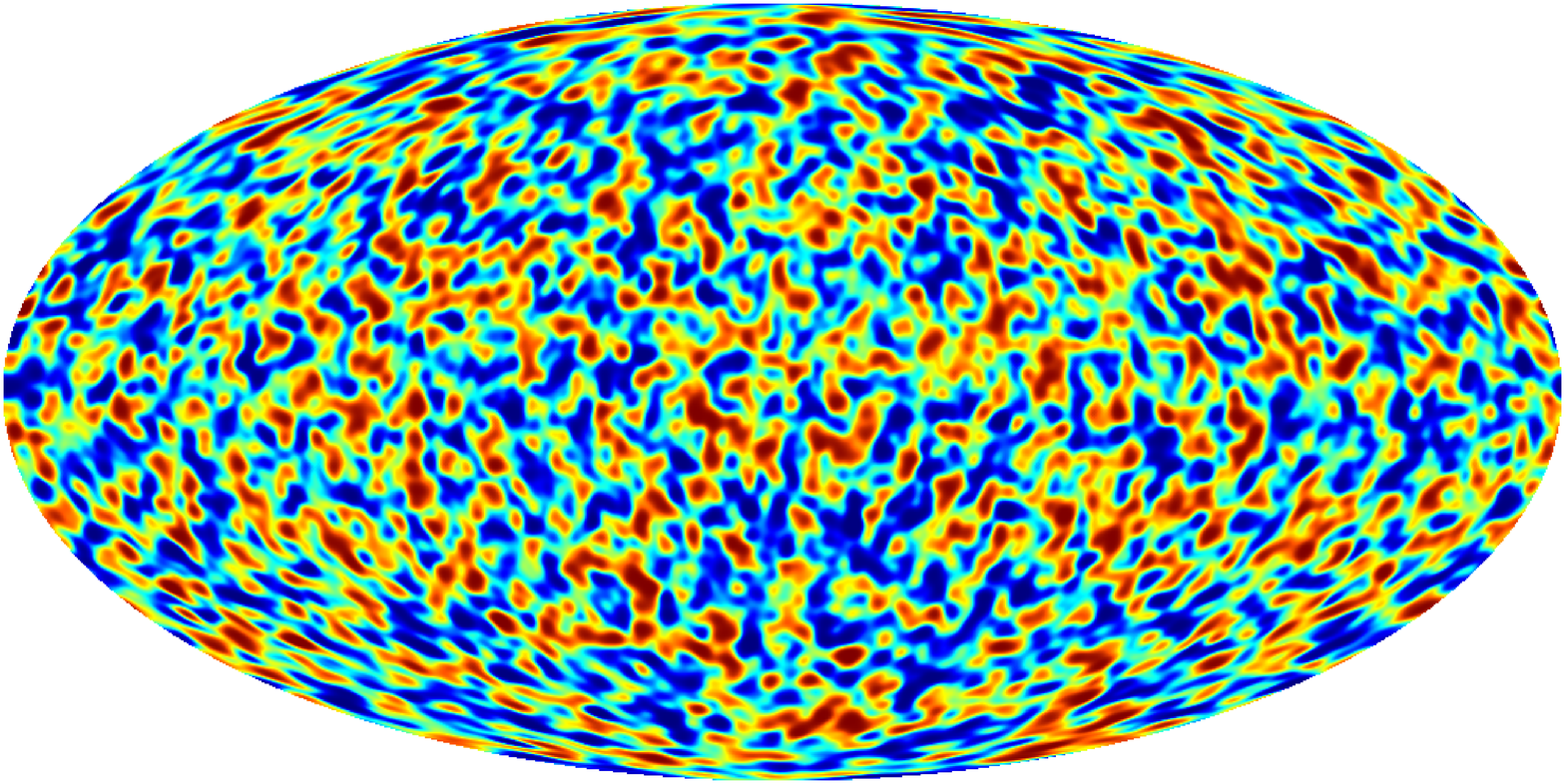}
\includegraphics[width=8.0cm]{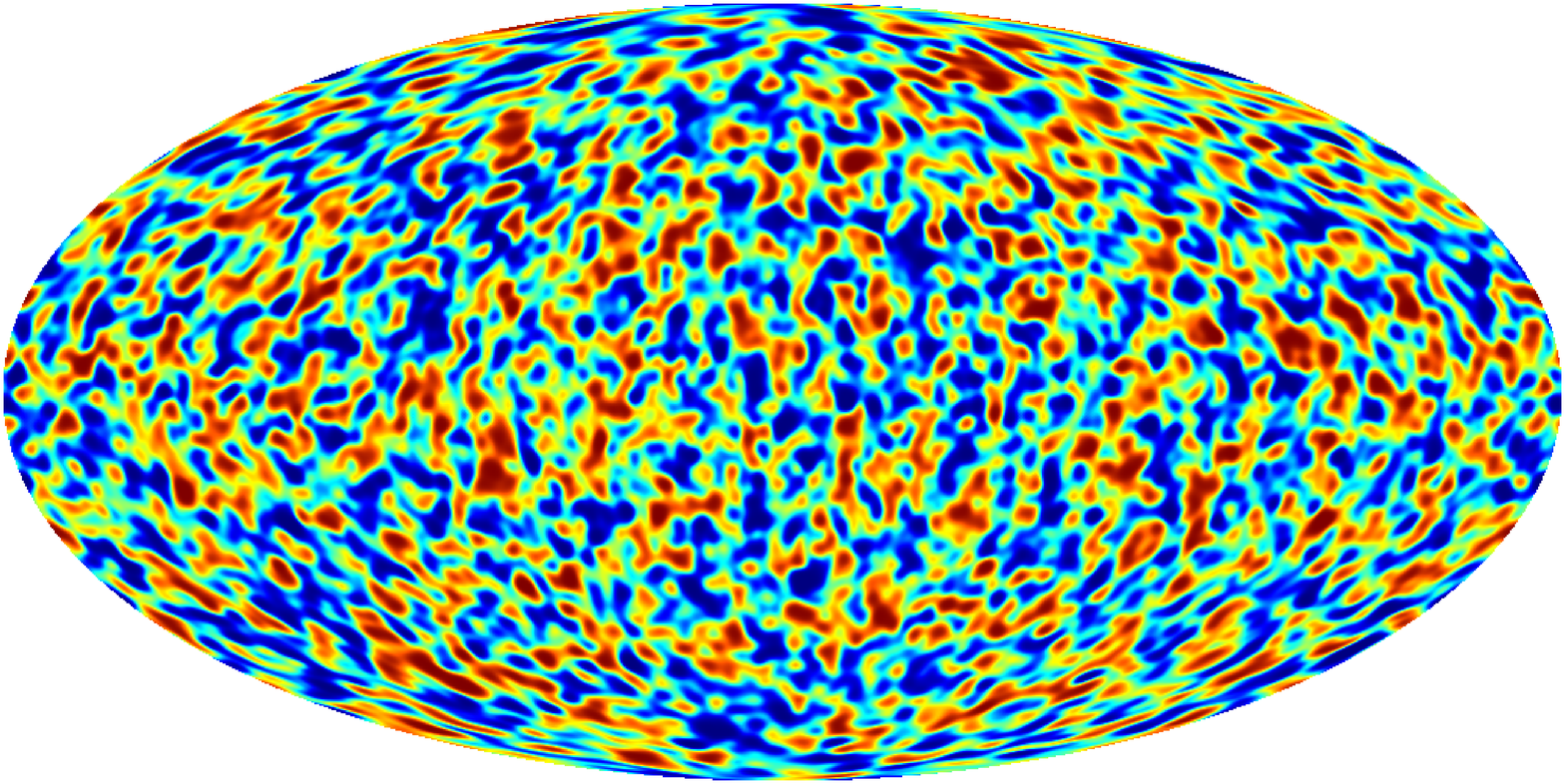}
\caption{Sky map, realized with HEALPix, showing a representative random realization of an isotropic distribution of objects in the sky (left panel) and a surrogate sky obtained from such a distribution (right panel). Color codes the fraction of objects falling in a pixel, while a Gaussian beam with $3^{\circ}$ dispersion is used.}
\label{fig-sky-iso-surr}
\end{figure}

\subsection{Uniform sky}\label{sec:unifsky}

In order to illustrate our procedure, we consider two examples over the whole sky: i) an isotropic distribution of objects in the sky (ISO toy model); ii) a distribution of objects where the 50\% is clustered around 10 hot spots randomly distributed in the sky, while the remaining objects are uniformly distributed (SRC toy model). 

\begin{figure}[!t]
\includegraphics[width=8.0cm]{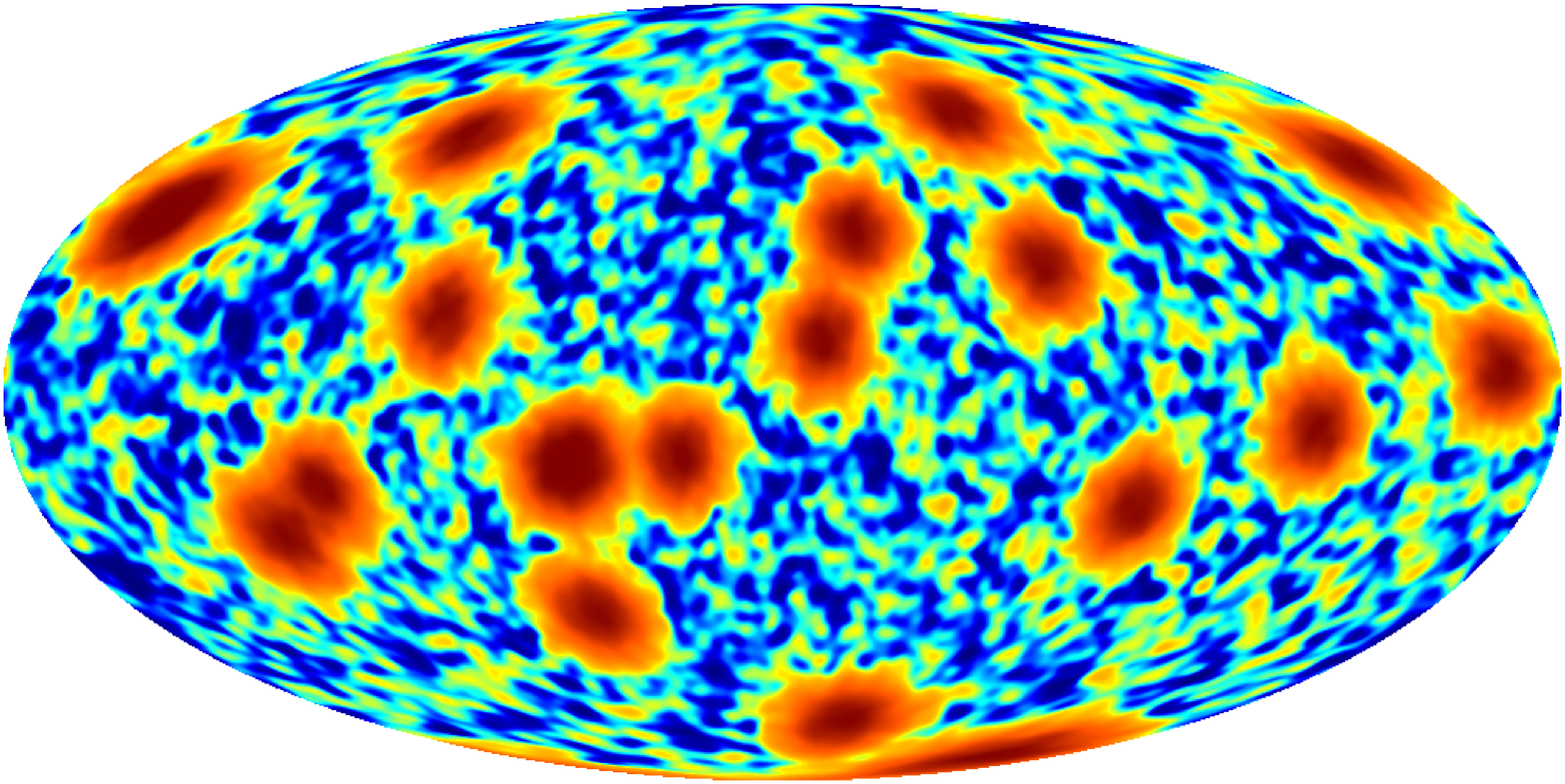}
\includegraphics[width=8.0cm]{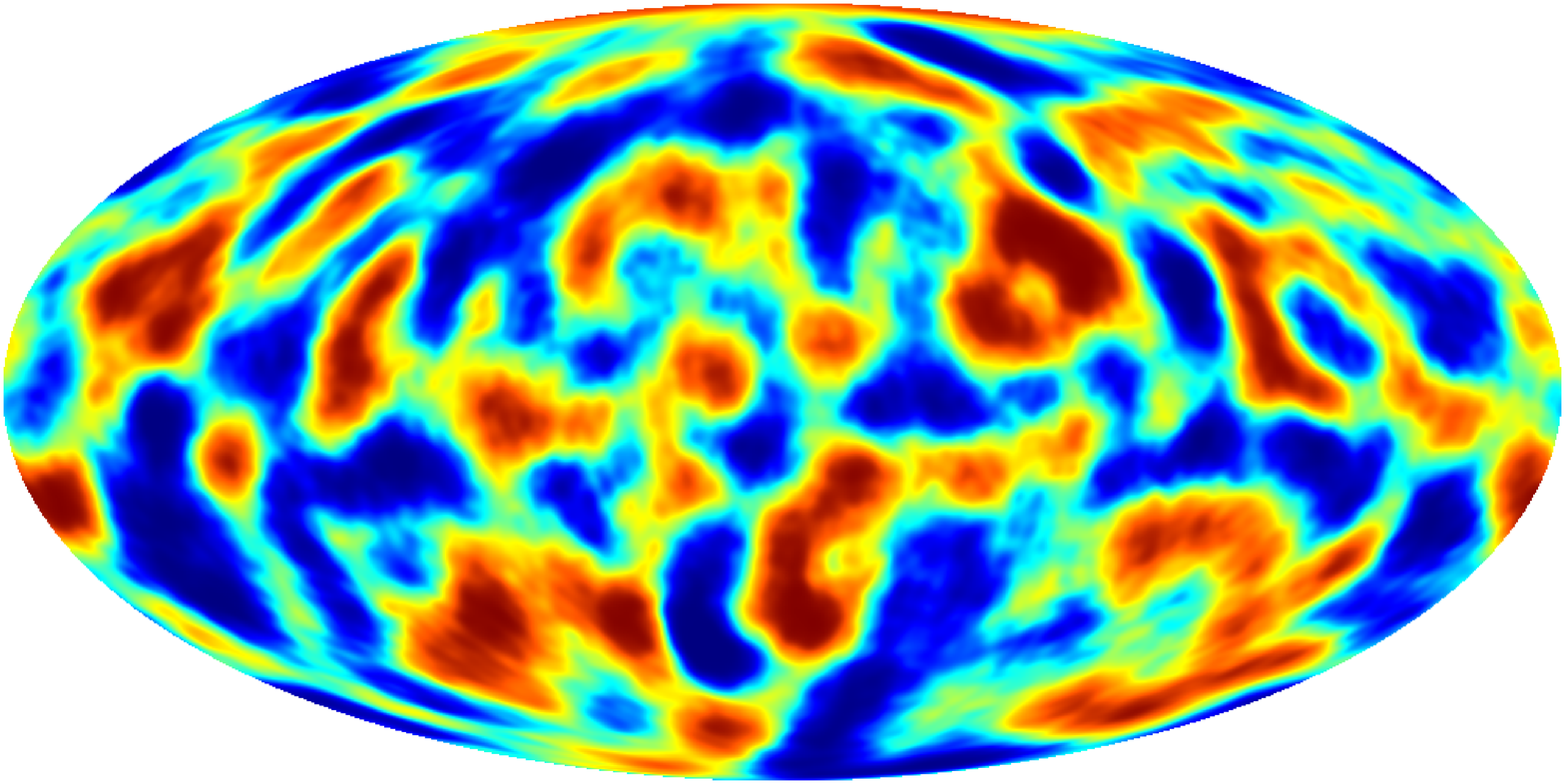}
\caption{Same as in Fig.\,\ref{fig-sky-iso-surr}, in the case of a sky where the 50\% is clustered around 10 hot spots randomly distributed in the sky, while the remaining objects are uniformly distributed.}
\label{fig-sky-src-surr}
\end{figure}

In Fig.\,\ref{fig-sky-iso-surr} we show the density map of the ISO toy model (left panel) and a realization of the corresponding surrogate map (right panel). In Fig.\,\ref{fig-sky-src-surr} we show the same maps in the case of the SRC toy model. Even from a simple by eye-inspection, it is evident that our procedure preserves the underlying structures of the original distribution of objects. 

\begin{figure}[!ht]
\centering
\includegraphics[width=12.0cm]{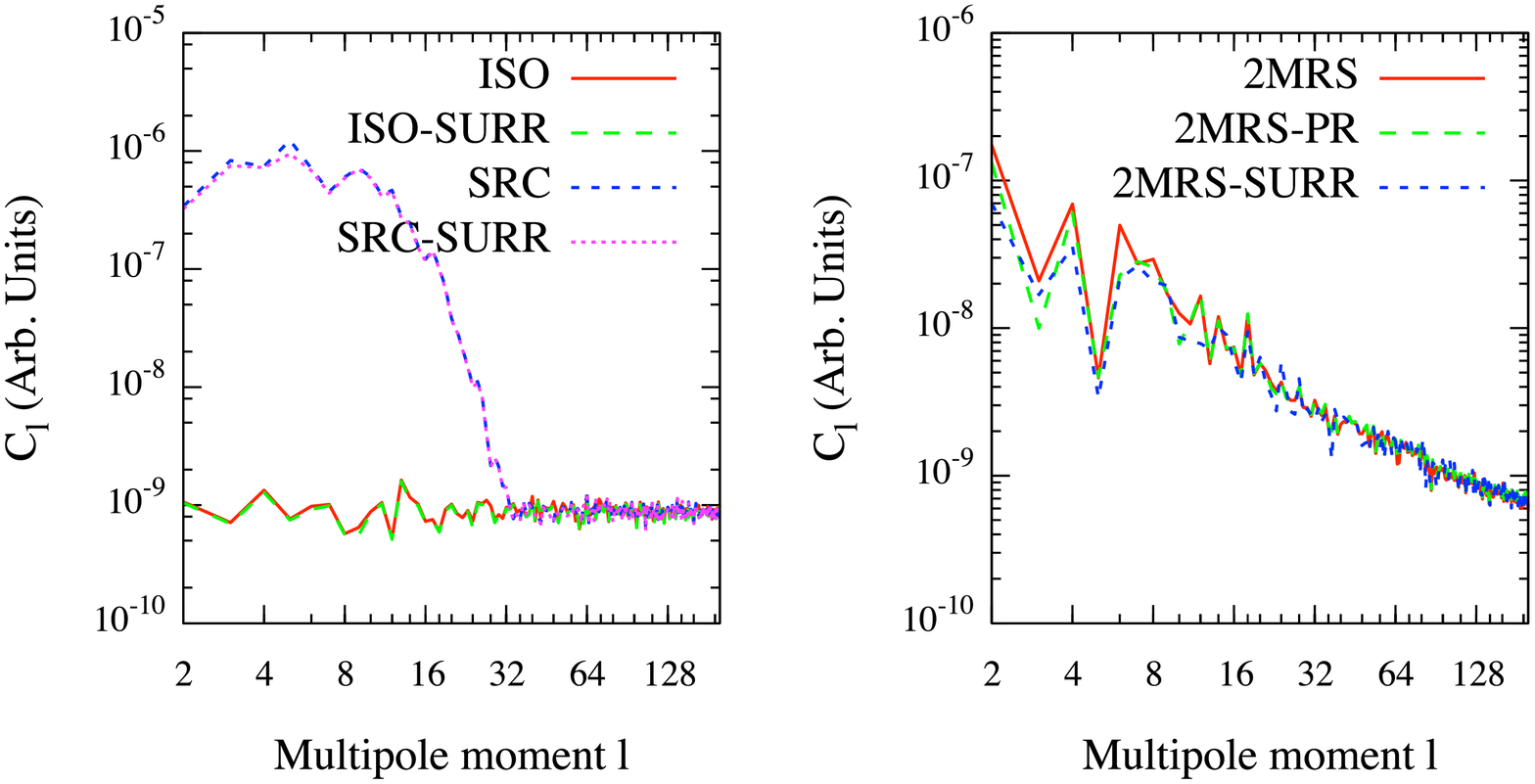}
\caption{\emph{Left panel:} angular power spectra of intensity maps shown in Fig.\,\ref{fig-sky-iso-surr} and Fig.\,\ref{fig-sky-src-surr}. Spectra for ISO (Fig.\,\ref{fig-sky-iso-surr}, left panel), ISO-SURR (Fig.\,\ref{fig-sky-iso-surr}, right panel), SRC (Fig.\,\ref{fig-sky-src-surr}, left panel) and SRC-SURR (Fig.\,\ref{fig-sky-src-surr}, right panel) are shown. \emph{Right panel:} Angular power spectra for 2MRS catalog of galaxies (2MRS), a phase-randomized surrogate (2MRS-PR) and a surrogate accounting for the mask (2MRS-SURR), corresponding respectively to upper, middle and lower panels of Fig.\,\ref{fig-sky-2mrs}.}
\label{fig-sky-spectra}
\end{figure}

Hence, we have estimated the angular power spectrum for each density map, in order to verify if it is preserved as well. In Fig.\,\ref{fig-sky-spectra} we show the angular power spectra corresponding to the density sky maps of our toy models. In any case our procedure correctly preserves the spectrum. In the case of the isotropic sky maps (see Fig.\,\ref{fig-sky-iso-surr}), the power spectrum is nearly flat, as expected for a typical ``white noise'' process, where objects are uniformly distributed on the sphere. In the case of the sky map of sources and background, the angular power spectrum has not a trivial shape. In any case, it is worth remarking that such results can be obtained for any shape of the power spectrum, and, by consequence, for any distribution of objects on the sphere, under the hypothesis of a uniform exposure.

\subsection{Sky including a mask}

At variance with the previous case, it may happen that the data are provided by observations of a sky with non-uniform coverage. In this case, a simple renormalisation can be applied to recover the case of the uniform coverage. Nevertheless, if certain regions of the sky have not been covered by a survey or simply they were not visible for a given reason, e.g. obstructed by a closer object, this procedure does not apply anymore. Therefore, the phase randomization method to be still applied requires some modifications. As an example, we consider the recently published 2MRS catalog of galaxies \cite{2MRScatalog2011}, where the region around the galactic plane is totally masked. The catalog is 97.6\% complete to a limiting magnitude of $K=11.75$ over 91\% of the sky, and contains almost 45,000 galaxies. The catalog provides a fair sample of the nearby Universe when the region corresponding to the Galactic band with $| b | \leq 5^\circ$ ($|b|\leq 8^\circ$ around the Galactic center) is excluded. Such a band defines a mask.

\begin{figure}[!t]
\centering\includegraphics[width=8.0cm]{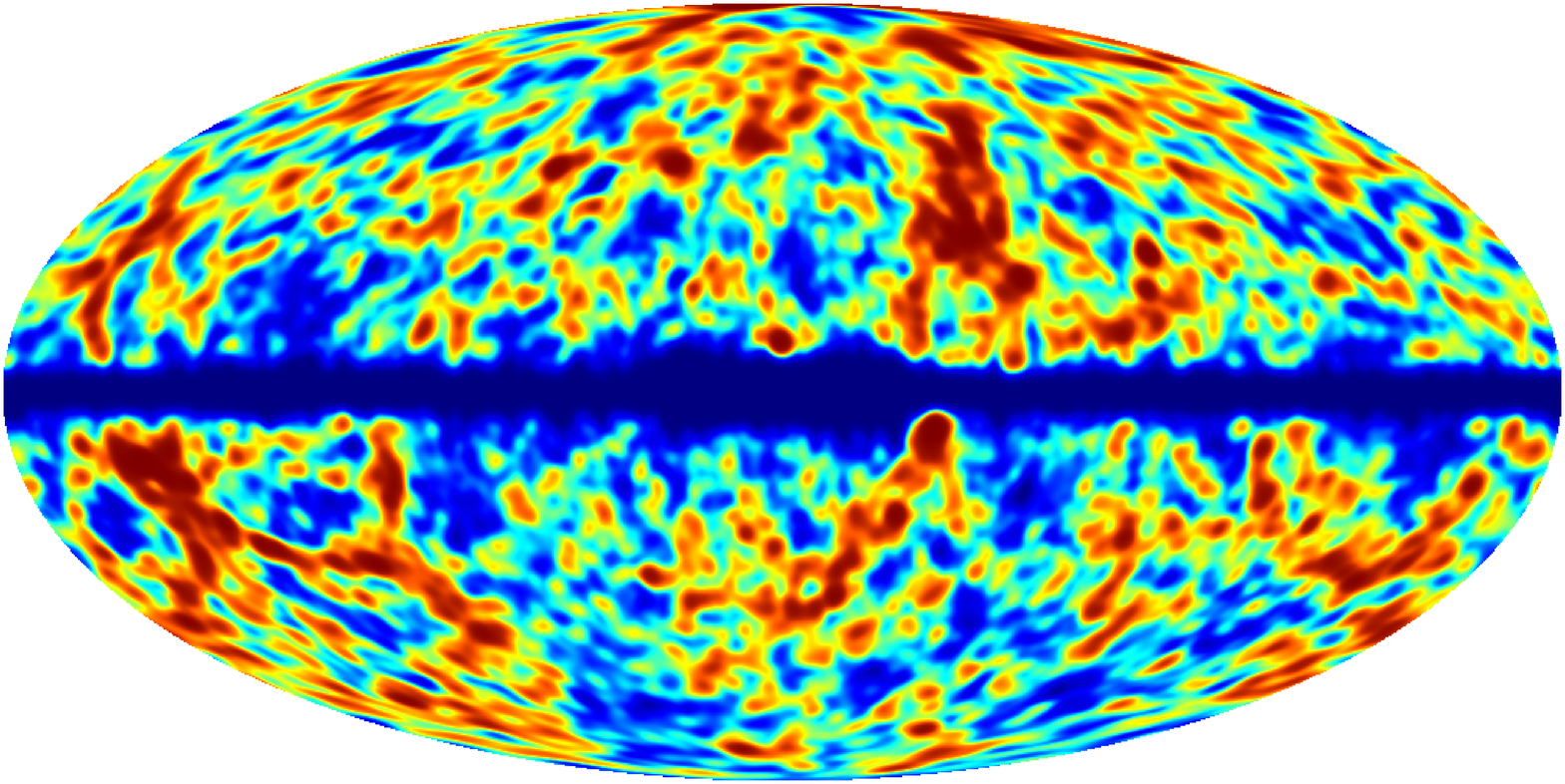}
\includegraphics[width=8.0cm]{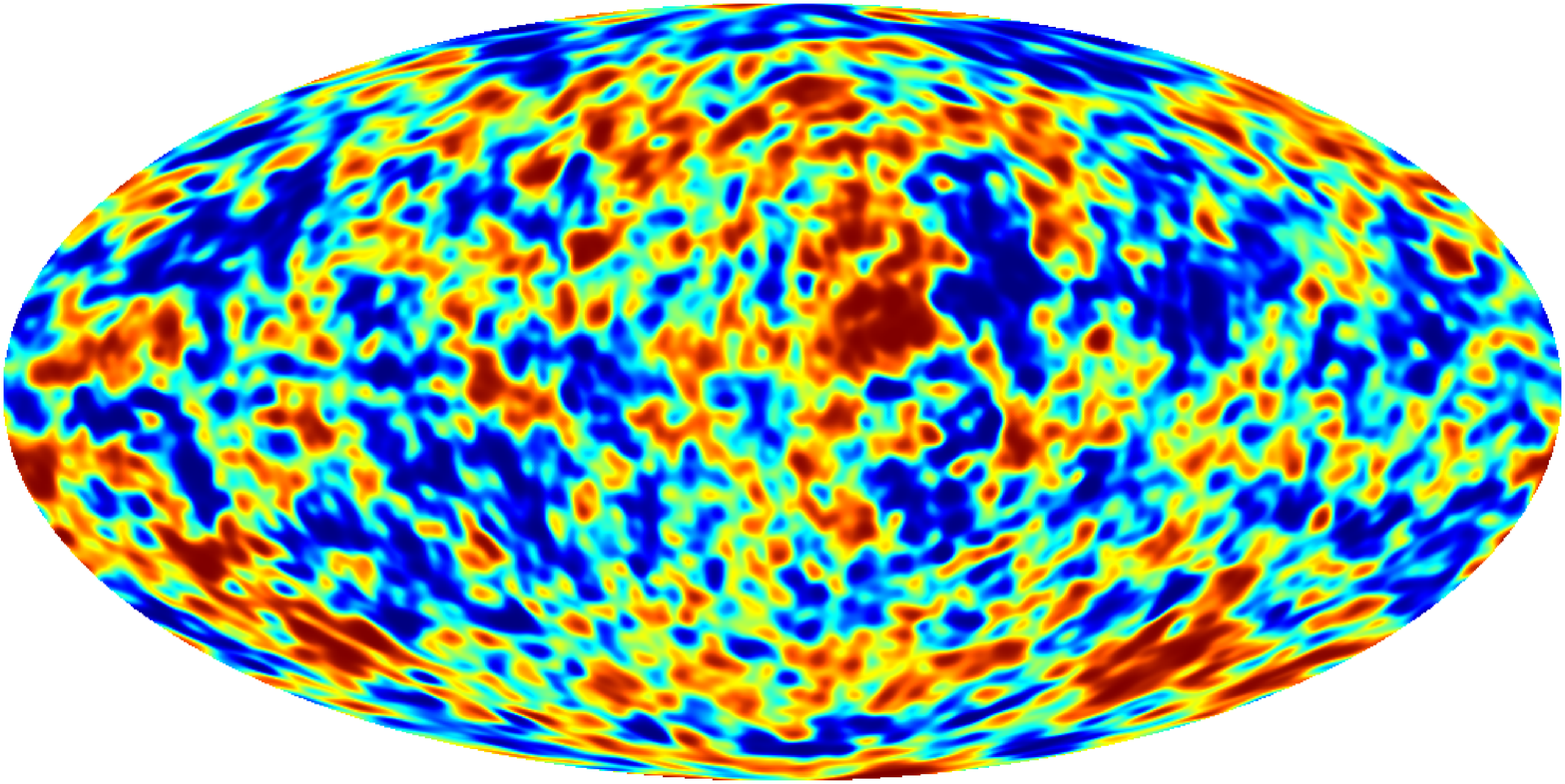}\\
\includegraphics[width=8.0cm]{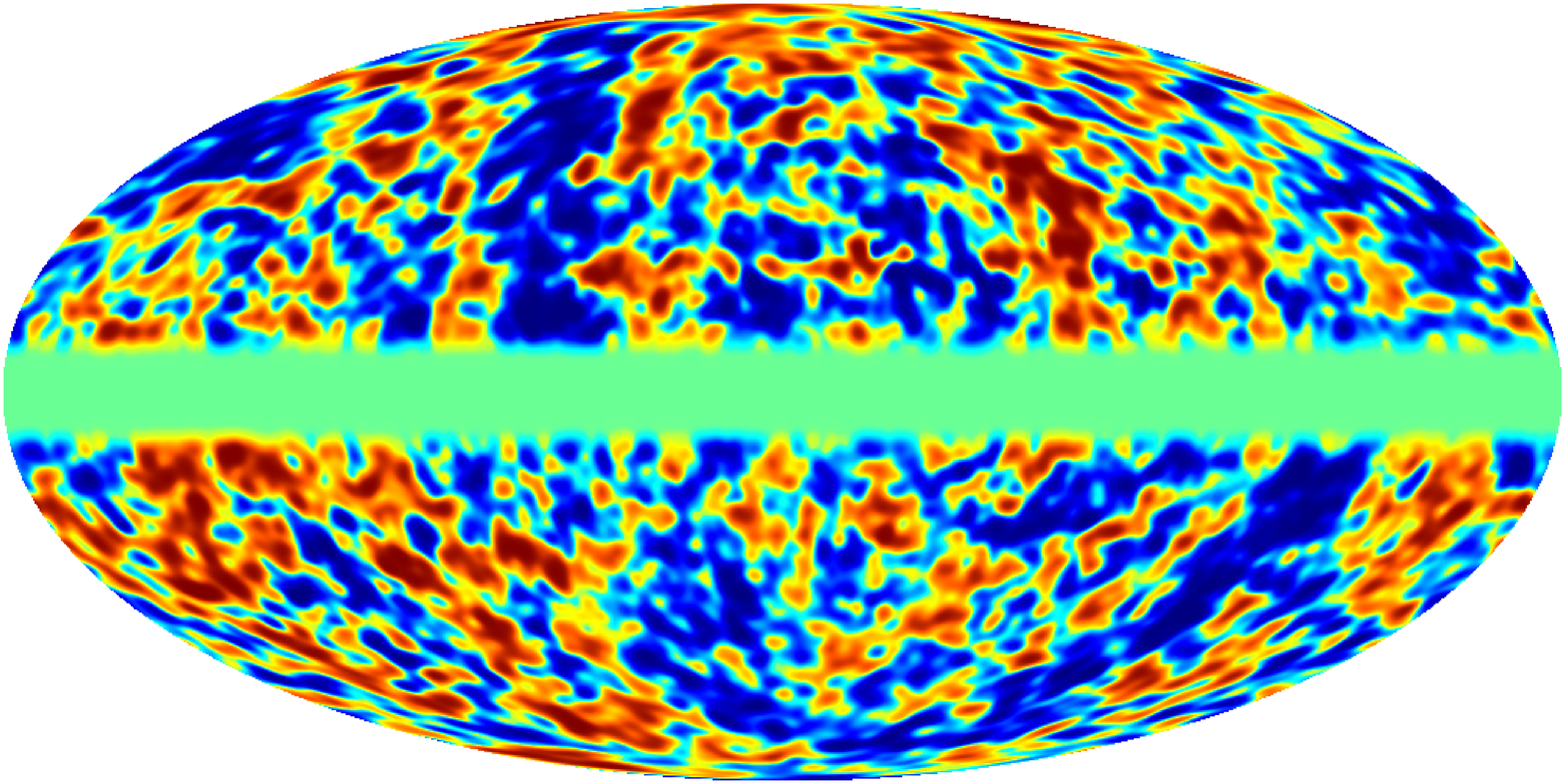}
\caption{Same as in Fig.\,\ref{fig-sky-iso-surr}, in the case of the 2MRS catalog of galaxies (upper panel). The simple phase-randomized surrogate of the catalog is shown in the middle panel: the information about the mask is lost. After taking into account such an information in the phase-randomization procedure, the correct surrogate map is obtained (lower panel).}
\label{fig-sky-2mrs}
\end{figure}

It is straightforward to show that the phase randomization preserves the power spectrum of the density map even in the case of an incomplete sky. However, it destroys the information about the mask, providing a surrogate map defined on the whole sky. Therefore, an additional step should be considered in such cases. 

Let $a_{lm}'$ indicate the phase-randomized multipole coefficients obtained from the expansion of the original map $f(\theta,\phi)$, and let $f'(\theta,\phi)$ indicate the resulting surrogate map. Hence, we assign a null weight to all pixels of the sphere covering the masked region and a weight proportional to $\Omega_{p}^{-1}$ to the remaining pixels, being $\Omega_{p}$ the solid angle covered by a single pixel. 

Finally, we estimate the new coefficients $a_{lm}''$ which take into account the weights assigned to pixels on the sphere, and we use again the spherical harmonics expansion defined by Eq.\,(\ref{def-sht}) to obtain the new surrogate map $f''(\theta,\phi)$, masked as the original one.



In Fig.\,\ref{fig-sky-2mrs} we show the density map of the 2MRS catalog of galaxies (upper panel). Hence, the simple application of the phase-randomization procedure, described at the beginning of the present section, produces a surrogate map over the whole sky (middle panel), where the information about the mask is lost. However, by using the extended version of our procedure, the masking effect is correctly taken into account, producing a surrogate map correctly masked (lower panel).

The angular power spectra corresponding to such three cases are shown in the right panel of Fig.\,\ref{fig-sky-spectra}: it is evident that both our procedures correctly preserve the spectra of the original skies.





\section{Discussion and conclusions}

We have presented a new method to generate random distributions of objects on the sphere with the same angular clustering of real distributions provided as input. Our method is based on the phase randomization of multipole coefficients corresponding to the spherical harmonics expansion of a map defined on the sphere. It preserves the angular power spectrum and, by consequence, the correlation function, because of the Wiener-Khinchin-Kolmogorov theorem. However, the method is model-independent and it does not require the direct estimation of the correlation function and the power spectrum of the original data, avoiding the intrinsic uncertainty on such estimates. Moreover, no fitting procedures are involved. We have shown that the method works on very different sky maps, and we have applied it to the distribution of galaxies in the 2MASS Redshift Survey as an application to real-world data.

The extension to the case of the spatial clustering is straightforward but requires much more computational efforts, and it did not represent the main goal of the present study. However, such a task can be achieved in a reasonable amount of computational time by means of the most advanced techniques for the numerical estimation of fast Fourier transforms (FFT) in the multi-dimensional space, based, for instance, on the pencil decomposition. 

Our method can be used in several applications. For instance, it can be used to generate sky maps of objects with a given angular power spectrum: the resulting maps can be used for several purposes, as to train new or existing methods for clustering detection of different types \cite{Peebles-1980,ave20092pt+,Hague09,Gaztanaga94Clusteringmethod,dedomenico2011MAF}.


Moreover, our method could be useful for estimating the overall expected errors in peculiar velocity field, where mock catalogs with the same features of a given real catalog are required, as shown in recent studies \cite{davis2011local, nusser2012new}.

Finally, the method is suitable to simulate realistic distributions of sources of ultra-high energy cosmic rays (UHECRs). In fact, the lack of information about the sources of UHECRs has motivated several investigations requiring the simulation of particles reflecting special type of clustering \cite{cuoco2008clustering,cuoco2009global,DeDomenicoICRC2011,takami2012sourcedensitylimit}, which can be easily provided by our procedure.


\section{Acknowledgements}
H. L. acknowledge the support of CAPES (Grant No. CAPES-PNPD 2940/2011).

\appendix

\newpage
\addcontentsline{toc}{section}{References} 
\begin{small}
\bibliographystyle{JHEP} 
\bibliography{draft}

\providecommand{\href}[2]{#2}\begingroup\raggedright\begin{thebibliography}{10}

\bibitem{Peebles-1980}
{P.J.E. Peebles}, {\em The Large Scale Structure of the Universe}.
\newblock Princeton Univ. Press, 1980.

\bibitem{Davis-1983}
{M. Davis and P.J.E. Peebles}, {\it A survey of galaxy redshifts vs the two
  point position and velocity correlations},  {\em Ap. J.} {\bf 267} (1983)
  465--482.

\bibitem{ramella19902ptcorfunc}
M.~Ramella, M.~J. Geller, and J.~Huchra, {\it {The two-point correlation
  function for groups of Galaxies in the center for Astrophysics redshift
  survey}},  {\em Astroph. J.} {\bf 353} (1990) 51--58.

\bibitem{davis1985evolution}
M.~Davis, G.~Efstathiou, C.~Frenk, and S.~White, {\it {The evolution of
  large-scale structure in a universe dominated by cold dark matter}},  {\em
  Astroph. J.} {\bf 292} (1985) 371--394.

\bibitem{springel2005simulations}
{V. Springel \it{et al}}, {\it {Simulations of the formation, evolution and
  clustering of galaxies and quasars}},  {\em Nature} {\bf 435} (2005),
  no.~7042 629--636.

\bibitem{Szalay-1993}
{S.D. Landy and A.S. Szalay}, {\it Bias and variance of angular correlation
  functions},  {\em Ap. J.} {\bf 412} (1993) 64--71.

\bibitem{Hamilton-1993}
{A.J.S. Hamilton}, {\it Toward better ways to measure the galaxy correlation
  function},  {\em Ap. J.} {\bf 417} (1993) 19--35.

\bibitem{ave20092pt+}
M.~Ave, L.~Cazon, J.~Cronin, J.~de~Mello~Neto, A.~Olinto, V.~Pavlidou,
  P.~Privitera, B.~Siffert, F.~Schmidt, and T.~Venters, {\it {The 2pt+: an
  enhanced 2 point correlation function}},  {\em J. Cosm. Astrop. Phys.} {\bf
  07} (2009) 023.

\bibitem{Hague09}
{J.D. Hague, B.R. Becker, M.S. Gold and J.A.J. Matthews}, {\it A three-point
  cosmic ray anisotropy method},  {\em J. Phys. G: Nucl. Part. Phys.} {\bf 36}
  (2009) 115203.

\bibitem{Gaztanaga94Clusteringmethod}
J.~M. Fry and E.~Gazta\~{n}aga, {\it {TO BEFOUND}},  {\em Astrophysics Journal}
  {\bf 415} (1994) 1.

\bibitem{dedomenico2011MAF}
{M. De Domenico, A. Insolia, H. Lyberis, M. Scuderi}, {\it {Multiscale
  autocorrelation function: a new approach to anisotropy studies}},  {\em J.
  Cosm. Astrop. Phys.} {\bf 03} (2011) 008,
  [\href{http://xxx.lanl.gov/abs/1001.1666}{{\tt 1001.1666}}].

\bibitem{cuoco2008clustering}
A.~Cuoco, S.~Hannestad, T.~Haugb{\o}lle, M.~Kachelrie{\ss}, and P.~Serpico,
  {\it {Clustering Properties of Ultra-High-Energy Cosmic Rays}},  {\em
  Astroph. J.} {\bf 676} (2008) 807--815.

\bibitem{cuoco2009global}
A.~Cuoco, S.~Hannestad, T.~Haugb{\o}lle, M.~Kachelrie{\ss}, and P.~Serpico,
  {\it {A global autocorrelation study after the first Auger data: impact on
  the number density of UHECR sources}},  {\em Astroph. J.} {\bf 702} (2009)
  825.

\bibitem{DeDomenicoICRC2011}
{M. De Domenico for the Pierre Auger Collaboration}, {\it {Bounds on the
  density sources of ultra high energy cosmic rays from the Pierre Auger
  Observatory data}}, .

\bibitem{takami2012sourcedensitylimit}
H.~Takami, S.~Inoue, and T.~Yamamoto, {\it {Propagation of Ultra-High Energy
  Cosmic Ray Nuclei in Cosmic Magnetic Fields and Implications for Anisitropy
  Measurements}},  {\em submitted to Astroph. J.} (2012)
  [\href{http://xxx.lanl.gov/abs/astro-ph/1202.2874v1}{{\tt
  astro-ph/1202.2874v1}}].

\bibitem{2MRScatalog2011}
{ J.P. Huchra \it{et al}}, {\it {The 2MASS Redshift Survey: description and
  data release}},  {\em ApJS} (2012) in press.

\bibitem{peacock2003large}
J.~Peacock, {\it {Large-scale surveys and cosmic structure}},  {\em ArXiv
  e-prints} (2003) [\href{http://xxx.lanl.gov/abs/0309240}{{\tt 0309240}}].

\bibitem{pietronero1987fractal}
L.~Pietronero, {\it {The fractal structure of the Universe: correlations of
  galaxies and clusters and the average mass density}},  {\em Physica A} {\bf
  144} (1987), no.~2-3 257--284.

\bibitem{coleman1992fractal}
P.~Coleman and L.~Pietronero, {\it {The fractal structure of the universe}},
  {\em Phys. Rep.} {\bf 213} (1992), no.~6 311--389.

\bibitem{martinez1999Universe}
V.~Mart{\'\i}nez, {\it {Is the Universe fractal?}},  {\em Science} {\bf 284}
  (1999), no.~5413 445.

\bibitem{joyce2000fractal}
M.~Joyce, P.~Anderson, M.~Montuori, L.~Pietronero, and F.~Labini, {\it {Fractal
  cosmology in an open universe}},  {\em Europh. Lett.} {\bf 50} (2000) 416.

\bibitem{guzzo1997redshift}
L.~Guzzo, M.~Strauss, K.~Fisher, R.~Giovanelli, and M.~Haynes, {\it
  {Redshift-space distortions and the real-space clustering of different galaxy
  types}},  {\em Astroph. J.} {\bf 489} (1997) 37.

\bibitem{peacock1999cosmological}
J.~Peacock, {\em {Cosmological physics}}.
\newblock Cambridge Univ Press, 1999.

\bibitem{peacock2001measurement}
{J.A. Peacock \it{et al}}, {\it {A measurement of the cosmological mass density
  from clustering in the 2dF Galaxy Redshift Survey}},  {\em Nature} {\bf 410}
  (2001), no.~6825 169--173.

\bibitem{hawkins20032df}
{E. Hawkins \it{et al}}, {\it {The 2dF Galaxy Redshift Survey: correlation
  functions, peculiar velocities and the matter density of the Universe}},
  {\em MNRAS} {\bf 346} (2003), no.~1 78--96.

\bibitem{fisher1994clustering}
K.~Fisher, M.~Davis, M.~Strauss, A.~Yahil, and J.~Huchra, {\it {Clustering in
  the 1.2-Jy IRAS galaxy redshift survey. II: Redshift distortions and $\xi$
  (rp, $\pi$)}},  {\em MNRAS} {\bf 267} (1994), no.~4 927--948.

\bibitem{ghigna1996deviations}
S.~Ghigna, S.~Bonometto, L.~Guzzo, R.~Giovanelli, M.~Haynes, A.~Klypin, and
  J.~Primack, {\it {Deviations from Hierarchical Clustering in Real and
  Redshift Space}},  {\em Astroph. J.} {\bf 463} (1996) 395.

\bibitem{saunders2000density}
{W. Saunders, W. \it{et al}}, {\it {Density and Velocity Fields from the PSCz
  Survey}},  in {\em Cosmic Flows Workshop}, vol.~201, p.~228, 2000.

\bibitem{HEALPix2005}
K.~M. Gorski, {\it {HEALPix: A Framework for High-Resolution Discretization and
  Fast Analysis of Data Distributed on the Sphere}},  {\em ApJ} {\bf 622}
  (2005) 759.

\bibitem{davis2011local}
M.~Davis, A.~Nusser, K.~Masters, C.~Springob, J.~Huchra, and G.~Lemson, {\it
  Local gravity versus local velocity: solutions for $\beta$ and non-linear
  bias},  {\em MNRAS} {\bf 413} (2011), no.~4 2906--2922.

\bibitem{nusser2012new}
A.~Nusser, E.~Branchini, and M.~Davis, {\it A new method for the determination
  of the growth rate from galaxy redshift surveys},  {\em Astroph. J.} {\bf
  744} (2012) 193.

\end{thebibliography}\endgroup
\end{small}

\end{document}